\documentstyle[preprint,aps]{revtex}
\tightenlines
\begin{document}
\draft
\title{Phenomenology for ${\vec p}{\vec p}\to pp\pi^{\circ}$}
\author{G. Ramachandran and P.N. Deepak\footnote[1]{E-mail: \tt pndeepak@yahoo.com}}
\address{Department of Studies in Physics, University of Mysore, 
Mysore 570 006, India}
\maketitle
\narrowtext
\begin{abstract}
We show that it is possible to empirically partition the differential 
cross section for $pp\to pp\pi^{\circ}$ into initial singlet and triplet 
differential cross sections in a model independent way which can be implemented 
using the existing technological capabilities at the PINTEX facility to study 
the reaction employing polarized beam or polarized
target or both simultaneously. 
\end{abstract}
\vskip 4.5in
\begin{center}
({\em Paper for presentation at the XIV DAE Symposium on High Energy Physics 
to be held at School of Physics, University of Hyderabad, Hyderabad, India 
from December 18-22, 2000})
\end{center}
\newpage
Considerable interest has been evinced in the study of pion production in 
$NN$ collisions using chiral perturbation theory \cite{chi} as 
high precision measurements \cite{tcs} of the total cross sections
for $pp\to pp\pi^{\circ}$ have become available.  With advances in storage ring 
technology \cite{mey}, the process has been studied more recently at the 
PINTEX facility \cite{expt} using a beam with polarization ${\bbox P}$ 
and target with polarization ${\bbox Q}$.    
To study the spin dependence, the measured
total cross section differences $\Delta\sigma _{T}$ and $\Delta\sigma _{L}$
have been analysed following Bilenky and Ryndin \cite{br} into
the singlet and triplet contributions $^{2s+1}\sigma _{m}$, where
$s$ denotes the initial channel spin and $m$ its projection along the beam 
direction.  Since the transferred momentum is large, the reaction is highly
sensitive to the short range $NN$ interaction and to the exchange mechanisms
involved therein.  Clearly an empirical study of the spin dependence at 
short distances necessitates experimental study of the differential 
cross section for ${\vec p}{\vec p}\to pp\pi^{\circ}$ as a function of the 
momentum transfer.

The purpose of this paper is to suggest a method to analyse measurements
of the differential cross section $d^{2}\sigma$ into singlet and triplet
differential cross sections $^{2s+1}d^{2}\sigma _{m}$.  This can be
implemented given the data, as functions of momentum transfer, on the 
unpolarized differential cross section $d^{2}\sigma _{0}$ together with 
measurements of $d^{2}\sigma$ employing ${\bbox P}$ alone, ${\bbox Q}$ 
alone and ${\bbox P}$ parallel to ${\bbox Q}$ along $\pm x,\pm y,\pm z$.

The suggested analysis is model independent and makes use of the irreducible
tensor formalism proposed recently \cite{nnpi}, following which the matrix
${\bbox M}$ in spin space may be expressed in the form
\begin{equation}
{\bbox M}=\sum _{s',s=0}^{1}
\sum _{\lambda =|s'-s|}^{s'+s}( S^{\lambda}(s',s)\cdot
M^{\lambda}(s',s)),
\label{M}
\end{equation}
where $s$ and $s'$ denote respectively the initial and final channel 
spins, the irreducible tensor operators $S^{\lambda}_{\mu}(s',s)$
of rank $\lambda$ are given by 
\begin{equation}
\label{S}
S^{\lambda}_{\mu}(s',s)=[s']\sum _{m}(-1)^{s-m}
C(s's\lambda;m'-m\mu)\ |s'm'\rangle\langle sm|
\end{equation}
following \cite{msv} and the irreducible tensor amplitudes 
$M^{\lambda}_{\mu}(s',s)$ are given by Eqs. (2) and (3) of \cite{nnpi}.

Using the same notations as in \cite{nnpi,msv} the differential 
cross section for ${\vec p}{\vec p}\to pp\pi^{\circ}$ is then given by
\begin{equation}
\label{dc}
d^{2}\sigma={\rm Tr}({\bbox M}\rho{\bbox M}^{\dagger})\ d^{3}p_{f}d\Omega=
{\rm Tr}({\bbox B}\rho),
\end{equation}
where 
\begin{equation}
\label{B}
{\bbox B}={\bbox M}^{\dagger}{\bbox M}\ d^{3}p_{f}d\Omega
\end{equation}      
is hermitian and the initial spin density matrix 
\begin{eqnarray}
\label{rho1}
\rho&=&{1\over 4}\left[1+\left({\bbox \sigma}_{1}\cdot{\bbox P}\right)\right]
\left[1+\left({\bbox \sigma}_{2}\cdot{\bbox Q}\right)\right]\\ 
\label{rho2}
&=& {1\over 4}\sum _{\alpha,\beta=0,x,y,z}\ \sigma _{1\alpha}\sigma _{2\beta}
\ P_{\alpha}Q_{\beta},
\end{eqnarray}
where $P_{0}=Q_{0}=1$ and $\sigma _{i0}$ for $i=1,2$ denote $2\times 2$ 
unit matrices.  Using Eq. (\ref{rho2}) in Eq. (\ref{dc}), the differential 
cross section may readily be expressed as
\begin{equation}
\label{dcB}
d^{2}\sigma=\sum _{\alpha,\beta}P_{\alpha}Q_{\beta}\ B_{\alpha\beta},
\end{equation} 
where 
\begin{equation}
\label{b}
B_{\alpha\beta}={1\over 4}{\rm Tr}\left[{\bbox B}\sigma _{1\alpha}
\sigma _{2\beta}\right].
\end{equation}
If both the beam and target are unpolarized, we readily identify the 
unpolarized differential cross section as 
\begin{equation}
\label{udc}
d^{2}\sigma _{_{0}}=B_{00}={1\over 4}{\rm Tr}({\bbox M}{\bbox M}^{\dagger})
d^{3}p_{f}d\Omega 
\end{equation} 
whereas 
\begin{equation}
d^{2}\sigma=B_{00}+P_{z}B_{z0}
\end{equation}
if the beam alone is polarized longitudinally,
\begin{equation}
d^{2}\sigma=B_{00}+Q_{z}B_{0z}
\end{equation}
if the target alone is polarized along the $z-$axis and
\begin{eqnarray}
\label{dc3}
d^{2}\Sigma _{\alpha}&=&{1\over 2}\left[
d^{2}\sigma _{\alpha}(\uparrow\uparrow)+d^{2}\sigma _{\alpha}
(\downarrow\downarrow)\right]\nonumber\\
&=& B_{00}+P_{\alpha}Q_{\alpha}\ B_{\alpha\alpha},\ \alpha=x,y,z
\end{eqnarray}
if $d^{2}\sigma _{\alpha}(\uparrow\uparrow)$ and $d^{2}\sigma _{\alpha}
(\downarrow\downarrow)$ denote respectively the differential cross 
sections when ${\bbox P}$ and ${\bbox Q}$ are polarized parallel to 
each other along $\pm \alpha$.  
Inverting (\ref{S}), using the 
unitarity of Clebsch Gordon coefficients after expressing the left 
hand side following \cite{msv} as  
\begin{equation}
\label{S1}
S^{\lambda}_{\mu}(s',s)={1\over 2}
[s']^{2}[s]\sum _{\lambda _{1},
\lambda _{2}=0}^{1}[\lambda _{1}][\lambda _{2}]
\ \left\{ \matrix{ 
{\textstyle {1\over 2}} &  {\textstyle {1\over 2}} & s' \cr
{\textstyle {1\over 2}} & {\textstyle {1\over 2}} & s  \cr
\lambda _{1} & \lambda _{2} & \lambda  }\right\}
\left(\sigma ^{\lambda _{1}}_{1}\otimes
\sigma ^{\lambda _{2}}_{2}\right)^{\lambda}_{\mu},
\end{equation} 
we obtain projection 
operators ${\bbox \pi}(s,m)=|sm\rangle\langle sm|$ given explicitly by
\begin{eqnarray}
{\bbox \pi}(1,1)&=&{1\over 4}\left[1+\sigma _{1z}+\sigma _{2z}+
\sigma _{1z}\sigma _{2z}
\right]\\ 
{\bbox \pi}(1,0)&=&{1\over 4}\left[1-2\sigma _{1z}\sigma _{2z}+\left(
{\bbox \sigma}_{1}\cdot {\bbox \sigma}_{2}\right)\right]\\
{\bbox \pi}(1,-1)&=&{1\over 4}\left[1-\sigma _{1z}-\sigma _{2z}
+\sigma _{1z}\sigma _{2z}
\right]
\end{eqnarray} 
which together add up to give the well known triplet projection operator
\begin{equation}
\label{T}
{\bbox T}=\sum _{m}{\bbox \pi}(1,m)={1\over 4}\left[3+\left(
{\bbox \sigma}_{1}\cdot {\bbox \sigma}_{2}
\right)\right],
\end{equation}
while 
\begin{equation}
{\bbox \pi}(0,0)={\bbox S}={1\over 4}\left[1-\left({\bbox \sigma}_{1}\cdot 
{\bbox \sigma}_{2}
\right)\right]
\end{equation}
denotes the well-known singlet projection operator.   Noting that 
$\sum _{s,m}{\bbox \pi}(s,m)=1$ and inserting the same between ${\bbox M}$ 
and ${\bbox M}^{\dagger}$ in Eq. (\ref{udc}), we readily obtain 
the partitioning of $d^{2}\sigma _{0}$ into $^{2s+1}d^{2}\sigma _{m}$ 
given explicitly by 
\begin{eqnarray}
\label{cdc1}
^{3}d^{2}\sigma _{_{+1}}&=&{1\over 4}\left(B_{00}+B_{z0}+B_{0z}+B_{zz}\right)\\
^{3}d^{2}\sigma _{_{0}}&=&{1\over 4}\left(B_{00}+B_{xx}+B_{yy}-B_{zz}\right)\\
^{3}d^{2}\sigma _{_{-1}}&=&{1\over 4}\left(B_{00}-B_{z0}-B_{0z}+B_{zz}\right)\\
\label{cdc4}
^{1}d^{2}\sigma _{_{0}}&=&{1\over 4}\left(B_{00}-B_{xx}-B_{yy}-B_{zz}\right)
\end{eqnarray}
in terms of six entities $B_{00},B_{z0},B_{0z},B_{xx},B_{yy}$ and 
$B_{zz}$ which are measurable experimentally, through (\ref{udc}) to 
(\ref{dc3}), using the existing technological capabilities at the 
PINTEX facility.

Although the reported experimental studies \cite{expt} have been carried out
at near threshold energies in order to study the lowest order partial wave
contributions to the final state, the method of analysis suggested above 
is clearly not specific to that energy region and may be adopted at higher 
energies as well.  In fact, it is applicable not only to $pp\to pp\pi^{\circ}$ 
but also to elastic $NN$ scattering, as well as to reactions initiated 
through a polarized nucleon beam incident on a polarized nucleon target.

We thank the Council of Scientific and Industrial Research (CSIR), 
India for support.

\end{document}